\documentclass[useAMS,usenatbib]{mn2e}
\usepackage{epsfig}
\def\ga{\mathrel{\raise0.35ex\hbox{$\scriptstyle >$}\kern-0.6em
\lower0.40ex\hbox{{$\scriptstyle \sim$}}}}
\def\la{\mathrel{\raise0.35ex\hbox{$\scriptstyle <$}\kern-0.6em
\lower0.40ex\hbox{{$\scriptstyle \sim$}}}}
\def\co{CO~{\it J}=1-0 }
\def\cotwo{CO~{\it J}=2-1 }
\def\cothree{CO~{\it J}=3-2 }
\def\hij{high-{\it J}~}
\def\loj{low-{\it J}~}
\def\arcs{\hbox{$^{\prime\prime}$}}

\title[A search for CO in HDF850.1]
{A broadband spectroscopic search for CO line emission in HDF850.1: 
the brightest submillimetre object in the \textit{Hubble Deep Field}
 North}
\author[Wagg et al. ]{
\parbox[t]{\textwidth}{
J. Wagg,$^{1,2}$\thanks{E-mail:jwagg@inaoep.mx} D.~H. Hughes,$^{1}$ I.~Aretxaga,$^{1}$ E.~L. Chapin,$^{3}$ J.~S. Dunlop,$^{4}$ \\
E. Gazta\~naga$^{5,1}$ and M. Devlin$^{6}$}
\vspace*{6pt}\\
$^{1}$ Instituto Nacional de Astrof\'{\i}sica, \'{O}ptica y Electr\'{o}nica 
    (INAOE), Apartado Postal 51 y 216, 72000 Puebla, Pue., Mexico\\
$^{2}$ Harvard-Smithsonian Center for Astrophysics, Cambridge, MA, USA, 02138\\
$^{3}$ Department of Physics and Astronomy, University of British Columbia, 
  6224 Agricultural, Vancouver, Canada, V6T 1Z1 \\
$^{4}$ SUPA\thanks{Scottish Universities Physics Alliance} 
   Institute for Astronomy, University of Edinburgh, Royal Observatory, 
   Blackford Hill, Edinburgh, EH9 3HJ, UK\\
$^{5}$ Institut d'Estudis Espacials de Catalunya, IEEC/CSIC, c/ Gran Capitan 
     2-4, 08034, Barcelona, Spain\\
$^{6}$  Department of Physics \& Astronomy, University of Pennsylvania, 
   209 South 33rd Street, Philadelphia, PA 19104-6396, USA\\
}

\begin{document}

\date{Accepted ....... Received ..............; in original form ......}
\pagerange{\pageref{firstpage}--\pageref{lastpage}} \pubyear{2006}

\maketitle

\label{firstpage}

\begin{abstract}

Using the 100-m Green Bank Telescope, we have conducted a cm-wavelength 
search for \co line emission towards the high-redshift, far-infrared luminous
object, HDF850.1 over the redshift interval 3.3~$\la z \la$~5.4. Despite 
the wealth of existing multi-wavelength observations, and the recent
 identification of a galaxy counterpart in 
deep $K^\prime$ band (2.2~$\mu$m) imaging, an unambiguous spectroscopic
 redshift has not yet been obtained 
for this object. A far-infrared-to-radio wavelength 
photometric redshift technique however, predicts a $\sim$90\% 
probability that the redshift is in the range,
 3.3~$\la z \la$~5.4 (equivalent to an observed redshifted \co emission line
 frequency, 26.5~$\ga \nu_{\rm obs} \ga$~18.0~GHz), making HDF850.1 a
 potential occupent of the `high-redshift tail' of submm selected galaxies.
 We have also conducted
 a search for \cotwo line emission over the narrower redshift range,
 3.9~$\la z \la$~4.3. Although we do not detect any CO line emission in this 
object, our limits to the CO line luminosity are in broad agreement with 
the median value measured in the current sample of high-redshift, 
submm selected objects detected in \hij CO line emission, but not sufficient
 to fully test the validity of the photometric redshift technique.  
\end{abstract}


\section{Introduction}

In recent years, blank-field extragalactic surveys at 
submillimetre/millimetre (hereafter submm) wavelengths have revealed a 
population of dusty galaxies undergoing vigorous star formation in the 
young Universe (e.g. Smail, Ivison \& Blain 1997; Hughes et al.\ 1998; 
Barger et al.\ 1998; Bertoldi et al.\ 2000; Scott et al.\ 2002; 
Borys et al.\ 2003;
 Greve et al.\ 2004; Laurent et al.\ 2005). Independent methods of
 redshift determination for these objects imply that the majority
 lie at, z~$\ga 2$ 
(Aretxaga et al.\ 2003, 2005; Chapman et al.\ 2003a, 2005). Thus their 
inferred star formation rates are in the range, 
100 to 1000~M$_{\odot}$~yr$^{-1}$, and arguably require large reservoirs
 of molecular gas for fuelling such high levels of sustained activity.
 Measuring the total molecular gas mass contained within the interstellar 
medium (ISM) of a high-redshift submm galaxy (hereafter SMG) is most 
effectively accomplished through observations of redshifted molecular 
CO line emission.

Over the past 15 years, observing CO line emission in 
high-redshift objects has become a powerful means of constraining the 
physical conditions of the gas within their molecular ISM
 (for an excellent review see Solomon \& Vanden~Bout 2005). 
The luminosity in the \co line is the optimal estimator of the 
total molecular gas mass yet, for practical reasons, in gas-rich 
objects at high-redshift (z~$\ga 2$), searches are normally first conducted
for emission from \hij  ({\it J$_{\rm upper}$}~$\ge$~2) CO line transitions,
 and subsequent searches for \co emission are carried out
 if the mm-wavelength searches for \hij CO lines are successful. This 
approach can bias the sample of objects detected in CO line emission to
 those with hotter, and denser gas. 

To date, \hij CO line emission has been detected in 14 SMGs 
(Frayer et al.\ 1998, 1999; Neri et al.\ 2003; Genzel et al.\ 2003; 
Downes \& Solomon 2003; Sheth et al.\ 2004; Kneib et al.\ 2005
Greve et al.\ 2005; Tacconi et al.\ 2006), while the \co line has
 been detected in only one of these (Hainline et al.\ 2006). 
In addition to the faintness of molecular emission lines in 
high-redshift SMGs, searches have been hindered by the limited
 spectral bandwidth of current mm-wavelength facilities, generally covering
$\sim$1700~km~s$^{-1}$ at 3~mm, while the typical SMG CO linewidth is
 $\sim$800~km~s$^{-1}$ FWHM. 
This narrow bandwidth can also be restrictive as galactic outflows
in many high-redshift SMGs may
lead to velocity offsets between the redshifts derived from 
Ly$\alpha$ and CO emission lines. In some SMGs this difference may be
 greater than 600~km~s$^{-1}$ 
 (e.g. Greve et al.\ 2005), due possibly to galactic outflows, or  
scattering of 
Ly$\alpha$ photons by dust. As CO emission-line frequencies are 
not expected to be biased with respect to the 
systematic redshift, broadband spectroscopic searches for CO line-emission 
should be a powerful means of obtaining redshifts for the SMG population.

Educated searches for mm-to-cm wavelength molecular CO line emission in 
luminous, dusty galaxies without redshifts, will become feasible in the
near future as wideband spectrometers are available on large mm-to-cm 
wavelength telescopes, for example the 100-m Green Bank Telescope 
(GBT\footnote{The Green Bank Telescope is a facility of the 
National Radio Astronomy Observatory, operated by Associated 
Universities, Inc. under a Cooperative Agreement with the 
National Science Foundation.}; Jewell \& Prestage 2004), or the 
50-m Large Millimeter Telescope 
(LMT\footnote{http://www.lmtgtm.org}). In order to obtain
redshift estimates, and to guide the frequency tunings of these 
spectroscopic searches for molecular line emission from SMGs, 
some groups have developed photometric redshift techniques which exploit the
far-infrared-to-radio wavelength correlation in star-forming galaxies 
(Helou et al.\ 1985),
 or adopt template far-infrared spectral energy distributions (SEDs) based on
 nearby galaxies (Carilli \& Yun 1999, 2000; Dunne, Clements
\& Eales 2000a; Rengarajan \& Takeuchi 2001;
Hughes et al.\ 2002; Aretxaga et al.\ 2003; Wiklind 2003; 
Hunt \& Maiolino 2005).
 This technique has the potential to provide redshift estimates for 
large samples of SMGs with individual accuracies, $\delta z \sim \pm$0.3,
 when photometric flux measurements of three or more far-infrared-to-radio 
wavelengths are available (Aretxaga et al.\ 2005).

The GBT is the only operational mm-to-cm wavelength telescope in the 
world with instruments that have both sufficient spectral line 
sensitivity and receiver bandwidth
 to conduct guided searches for CO line emission in SMGs at redshifts
 $z \ga 0.9$. This lower redshift limit is set by the current
 GBT frequency limit of 60~GHz and the \co line rest frequency of 
 115.2712~GHz. 
 Given this restriction,
an excellent candidate for conducting a blind search for CO line
 emission is HDF850.1 (Hughes et al.\ 1998), one of the most well
 studied SMGs, and the brightest 850~$\mu$m source in the 
confusion limited JCMT/SCUBA survey of the northern 
\textit{Hubble Deep Field}. 
Due partly to the extreme optical faintness ($K \simeq$~23.5, 
$I- K > 5.2$) of the gravitationally lensed galaxy counterpart 
to HDF850.1 (Dunlop et al.\ 2004), the redshift of this object has proven
 elusive. A wealth of deep rest-frame far-infrared-to-radio wavelength
 observations of 
HDF850.1 provide the basis for a photometric redshift
 z$=$4.1$\pm$0.5 (Yun \& Carilli 2002; Aretxaga et al.\ 2003). In principle,
HDF850.1 presents an ideal target for the GBT with which to test the accuracy
 of our photometric redshift technique, and thus has motivated a GBT 
search for \co and \cotwo line emission over the redshift interval, 
3.3~$\la z \la$~5.4.

We present the results of a GBT search for \co line emission in HDF850.1 
over the redshift interval, 3.3~$\la z \la$~5.4, and a search for 
\cotwo line emission over the narrower 
redshift interval, 3.9~$\la z \la$~4.3. Throughout this work, we adopt
 the following $\Lambda$-dominated cosmological parameters:
 $H_0 = 70$~km~s$^{-1}$~Mpc$^{-1}$, $\Omega_\Lambda = 0.7$, $\Omega_m = 0.3$
(Spergel et al.\ 2003, 2006).

\section{Observations and Data Reduction}

The far-infrared-to-radio wavelength photometric redshift estimate of
Aretxaga et al.\ (2003) implies an 86 to 90\% probability that 
HDF850.1 has a redshift in the range, 3.3~$\le z \le$~5.4. Over this 
redshift interval, the 115.2712~GHz \co line is redshifted into the
18.0 to 26.5~GHz frequency window of the K-band receiver on the GBT. 
Motivated by this prediction, we have obtained a complete K-band 
spectrum of HDF850.1 in order to search for redshifted \co line emission.

\begin{figure}
\epsfig{file=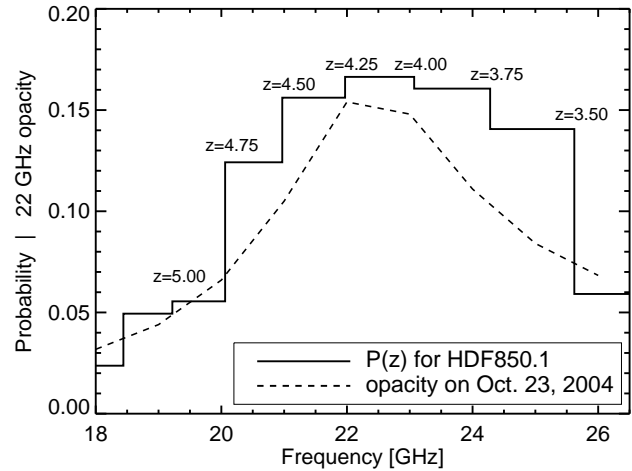,width=\hsize}

\caption[Redshift probability distribution as a function of redshifted
CO J=1-0 line frequency.]{The most recent  
submm-to-radio-wavelength photometric redshift estimate
for HDF850.1, plotted as a function of redshifted (115.2712~GHz) \co line 
frequency (\textit{solid line}). The \textit{dashed line} shows an example of 
the opacity across the K-band due to the atmospheric water vapour line at 
$\sim$22~GHz.  
}
\label{fig:tunings}
\end{figure}

Observations in nod mode were carried out with the GBT K-band 
receiver during October 2004 and May 2005. The position center 
adopted for the on-source beam was that of the mm-wavelength 
counterpart detected with the Plateau de Bure Interferometer (PdBI) by 
Downes et al.\ (1999). All of the K-band 
observations were conducted under reasonably dry 
conditions, with an average 22~GHz zenith opacity, 
$\tau_{\rm 22GHz} \sim 0.09$. The nearby quasar 3C295 was used for pointing 
purposes, as well as baseline and flux calibration throughout the K-band 
observations. 
 To correct for slowly varying, large-scale spectral baseline features, 
the observations were made by alternately nodding two beams separated 
by 178.8\arcs, between the source and blank sky. The GBT spectrometer 
allows a maximum instantaneous frequency coverage of 800~MHz bandwidth 
in each of 4 independent quadrants. For the observations presented here, 
one pair of quadrants was used to measure a $\sim$1.5~GHz wide spectra
 on the source, while the other pair measured blank sky in the off-beam. 
A total of 6 tunings (or sequences) were therefore used to cover the 
entire K-band window. 
Each spectral channel was 0.39~MHz wide so that the velocity resolution 
varied from $\sim$4.4~km~s$^{-1}$ to $\sim$6.5~km~s$^{-1}$ across the 
band. A total of 28.2~hours of integration time was devoted to the HDF850.1
K-band spectrum. The time spent on each $\sim$1.5~GHz tuning sequence was  
varied to compensate for the increased opacity towards the center of the 
band, due to the 22~GHz atmospheric water vapour line. The goal was to
obtain a spectrum with uniform noise across the K-band. Overheads such 
as pointing, focusing, acquisition of baseline calibration spectra, and
 follow-up of potential CO line detections, amounted
to an additional factor of 2 increase in the observing time.

In December 2005, a search was also conducted for \cotwo emission using the 
GBT Q-band receiver (40 to 48~GHz), over the redshift interval, 
3.91~$\la z \la$~4.25. As in the case of the K-band observations, 
the nod observing mode was adopted, while the spectrometer was 
set up in wide bandwidth, low resolution mode. The velocity resolution
varied from $\sim$2.5~km~s$^{-1}$ to $\sim$2.7~km~s$^{-1}$ across our 
Q-band spectrum. As there is a dearth of bright, compact calibration 
sources at these higher frequencies, the primary flux calibrator was
 3C286, while the nearby quasar 1153+495 ($\sim$17$^o$ separation)
 was used for pointing.

\begin{figure*}
\epsfig{file=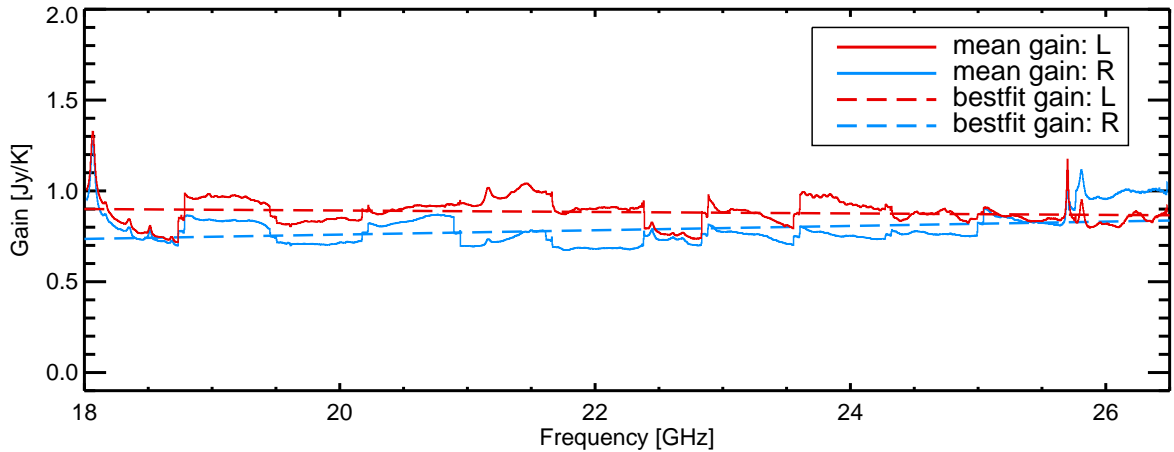,width=6.5in}
 \caption[Average gain across the GBT K-band window during our observations.]
{The GBT K-band gain (in Jy/K) for the left and right polarizations, averaged
over all observations of the calibrator, 3C295 (\textit{solid lines}). Also
plotted as \textit{dashed lines} are the bestfit linear regression curves
 to the data. These fits are then used to calibrate the combined HDF850.1
spectra. 
}
\label{fig:kband_gain}
\end{figure*}

Obtaining quality spectra with the GBT Q-band receiver generally requires 
low wind speeds ($\le$3~m~s$^{-1}$), and an extremely dry, stable 
atmosphere. Only on a single night in December 2005, was data obtained 
under just such 
conditions, with a median $\tau_{\rm 44GHz} \sim 0.1$ and negligible wind
speeds. Despite acceptable Q-band weather conditions during this, and 
possibly one other observing shift, only the left polarization Q-band 
spectra is included
in our analysis, as the majority of the right polarization spectra suffer
 from a severe baseline ripple of unknown origin (see \S\ref{sec:baselines}).

Both the K-band and Q-band spectra were reduced using the 
\textit{gbtidl}\footnote{http://gbtidl.sourceforge.net} data analysis package.
 For a series of consecutive scans, 
a co-added spectrum is produced following the standard procedures described
by Vanden Bout, Solomon \& Maddalena (2004), which are only outlined here 
for completeness.

 For simplicity, we will consider the spectrum in only one quadrant, and 
one polarization from each of the two beams in the discussion that follows.
 Let us refer to these spectra
as $B_1(\nu)$ and $B_2(\nu)$, where one of the two beams is always pointed 
`ON' the source, while the `OFF' beam is observing blank sky separated from 
the source by 178.8\arcs in azimuth. A single scan is the time that one beam
 spends on-source before nodding to the off-source position, when the 
other beam 
is nodded onto the source. For our observations, a scan duration of 1 minute,
during which one beam is on-source before nodding to the off-source position,
was adopted. This scan duration is chosen so as to minimize the frequency 
dependent variation in the sky brightness temperature between successive
 scans, while also spending less time on overheads such as nodding. 
Assuming that $B_1(\nu)$ begins
on-source at scan $i$, then a series of normalized spectra are produced by
subtraction of the off-source scan from the on-source scan
 (note that from this point on, the $\nu$ dependance will not be made 
explicit):\\ 

%\begin{center}
$\left (B_{\rm 1-ON}^{(i)}-B_{\rm 1-OFF}^{(i+1)} \right )/B_{\rm 1-OFF}^{(i+1)},$\\

$\left (B_{\rm 2-ON}^{(i+1)}-B_{\rm 2-OFF}^{(i)} \right )/B_{\rm 2-OFF}^{(i)},$\\

$\left (B_{\rm 1-ON}^{(i+2)}-B_{\rm 1-OFF}^{(i+3)} \right)/B_{\rm 1-OFF}^{(i+3)},$\\

$\left (B_{\rm 2-ON}^{(i+3)}-B_{\rm 2-OFF}^{(i+2)} \right)/B_{\rm 2-OFF}^{(i+2)},$\\

$\left (B_{\rm 1-ON}^{(i+4)}-B_{\rm 1-OFF}^{(i+5)} \right)/B_{\rm 1-OFF}^{(i+5)},$\\

$\left (B_{\rm 2-ON}^{(i+5)}-B_{\rm 2-OFF}^{(i+4)} \right)/B_{\rm 2-OFF}^{(i+4)},$\\

etc....\\ 
%\end{center} 

\noindent  It is important that each normalized spectrum be created
from the `ON'-`OFF' scan pair that travel down the same signal path 
(ie. the signal from a single beam and polarization), so that any 
path-dependent sources of baseline structure may be properly 
subtracted. 
An average of these normalized spectra is then calculated, with each
spectrum weighted by the inverse square of the system temperature, 
$T_{\rm sys}^{-2}$. The units of this mean spectrum are then calibrated from 
[K] to [Jy] using the best-fit gain curves shown for the K-band observations
 in Figure~\ref{fig:kband_gain}. The gain curves were 
derived from observations of objects with known (and non-variable) radio 
flux densities (3C295 for the K-band spectrum, and 3C286 for the
 Q-band spectrum).

The left and right polarization spectra are analyzed separately so that 
any potential detection of CO
 line emission would have to be confirmed in both polarizations. A large
fraction of the
 data ($\sim$53\%) were considered unusable due to various forms of 
spectral baseline 
irregularities and contamination which could not be removed 
reliably during the data 
reduction process (see \S\ref{sec:baselines}). We attempted to
 use spectra of the bright pointing sources to correct the 
 baseline shapes in these spectra (Vanden Bout, Solomon \& Maddalena 2004),
however we found that this did not improve our results. After removal of 
the poor quality data, the total on-source integration time devoted
 to the final K-band 
spectra is 12.3 hours in the left polarization spectra, and 13.9 hours in 
the right polarization spectra (see Table~\ref{tab:obstab}).

\begin{table}
\caption{Summary of K-band observations of HDF850.1. The total integration
time, $t_{\rm int}$, is based on the data which contributes to the final, 
left (L) and right (R) polarizaion spectra.  
\label{tab:obstab}}
\begin{center}
\begin{tabular}{cccc} \hline \hline
Seq.\# & $\nu_1 - \nu_2$[GHz] & $t_{\rm int}$ (L/R) [hrs] & $\bar{\sigma}$ (L/R) [mJy] \\
\hline 
K1 & $22.10-23.62$ & 2.7 / 3.1 & 0.58 / 0.69 \\
K2 & $23.54-25.06$ & 2.2 / 1.8 & 0.51 / 0.61 \\
K3 & $20.88-22.40$ & 2.1 / 3.1 & 0.50 / 0.52 \\
K4 & $19.44-20.96$ & 1.8 / 2.6 & 0.60 / 0.46 \\
K5 & $24.98-26.50$ & 1.5 / 1.2 & 0.98 / 0.85 \\
K6 & $18.00-19.52$ & 1.9 / 2.1 & 0.77 / 0.72 \\
\hline
\end{tabular}
\end{center}
\noindent
\end{table}

\subsection{Spectral Baselines}
\label{sec:baselines}

A major obstacle faced by searches for faint, broad emission lines, is 
the presence of variations in the spectral baseline shape, with 
a characteristic scale similar to the expected CO linewidths.
These features are not uncommon in mm/cm wavelength spectra obtained 
with single-dish instruments, and can easily be mistaken for 
 detections of weak, broad emission lines. There are a number of 
instrumental, as well as atmospheric effects which may produce spectral
baseline artefacts. Here, we summarize a few such artefacts which have
 been identified in our GBT (K and Q-band) spectra of HDF850.1: 

\begin{figure}
\epsfig{file=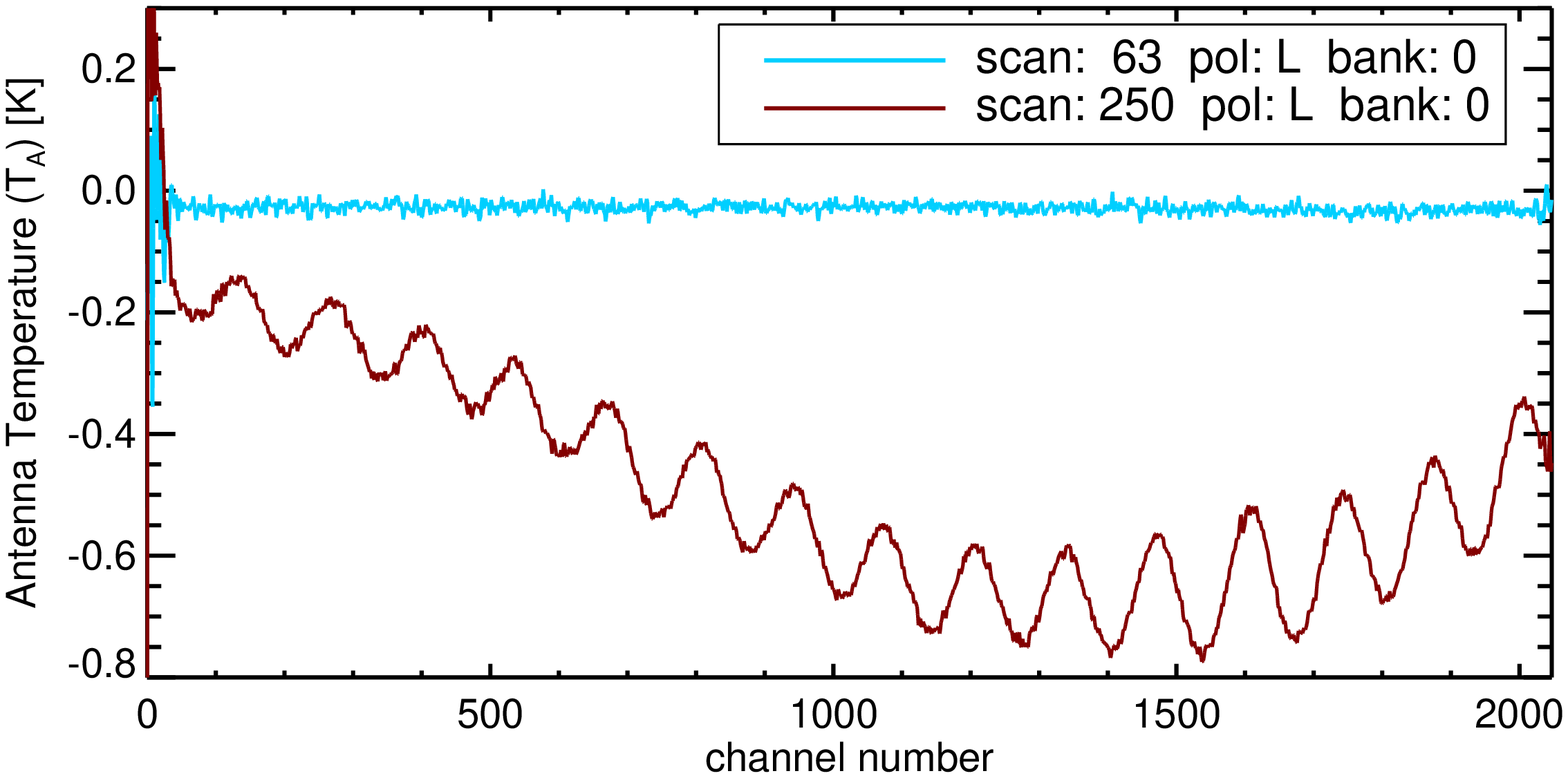,width=\hsize}
\epsfig{file=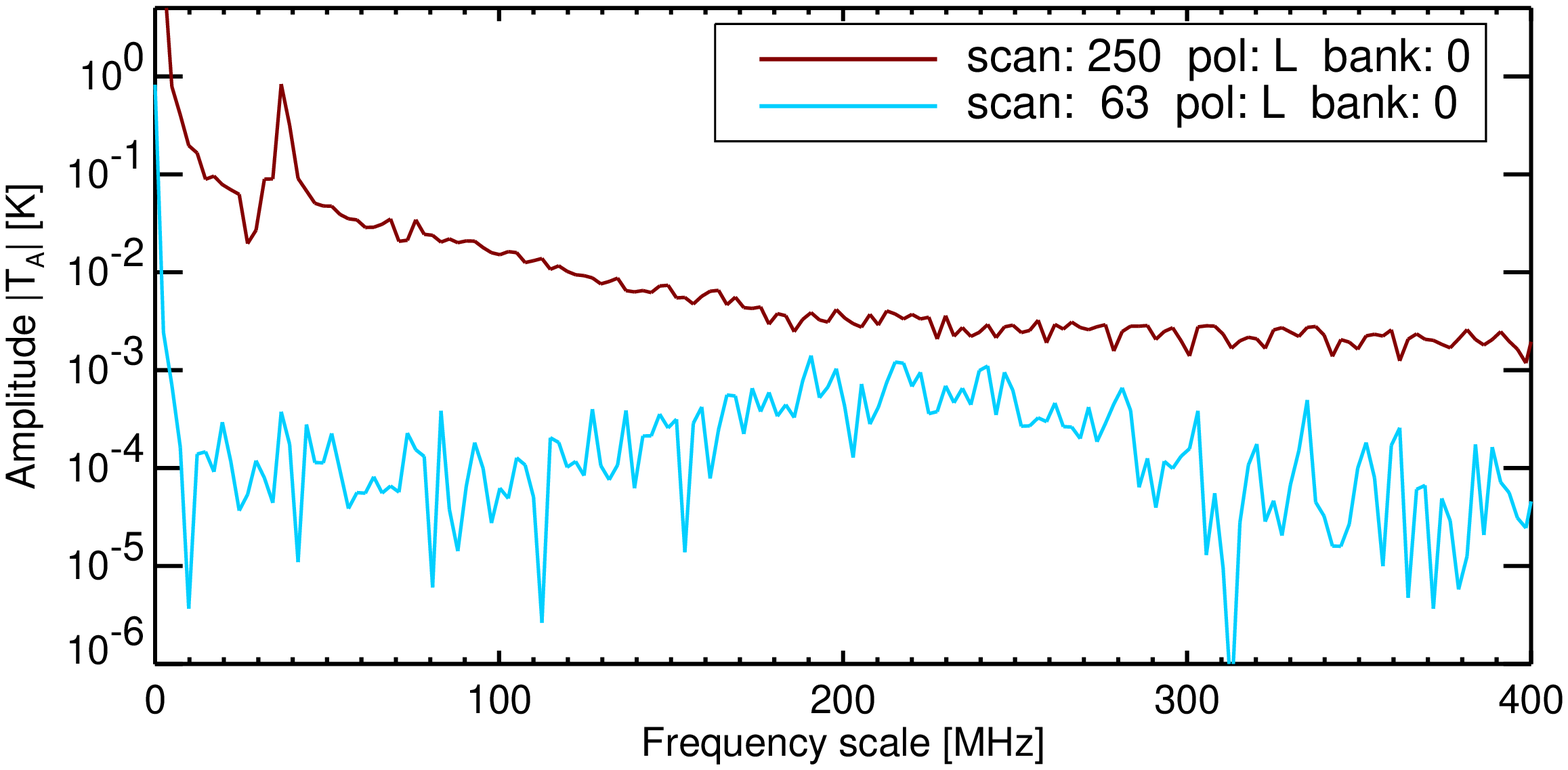,width=\hsize}
\caption[Example of the 50~MHz spectral baseline ripple and the corresponding 
power spectrum.]{\textbf{a)} An example of the $\sim$50~MHz ripple which 
infected much of the left polarization, K-band spectra during the May, 2005 
observing run. For comparison, we plot a spectrum of the same 
frequency tuning and polarization obtained on a different date.
 \textbf{b)} The power spectra of the two spectra, 
exemplifying the prominant spike on scales of $\sim$50~MHz in one of
 the two. This excess in power is likely due to temperature variations in the 
spectrometer room, which can result in the creation of standing waves 
in some of the signal connectors. 
}
\label{fig:rip50}
\end{figure} 

\begin{figure*}
\epsfig{file=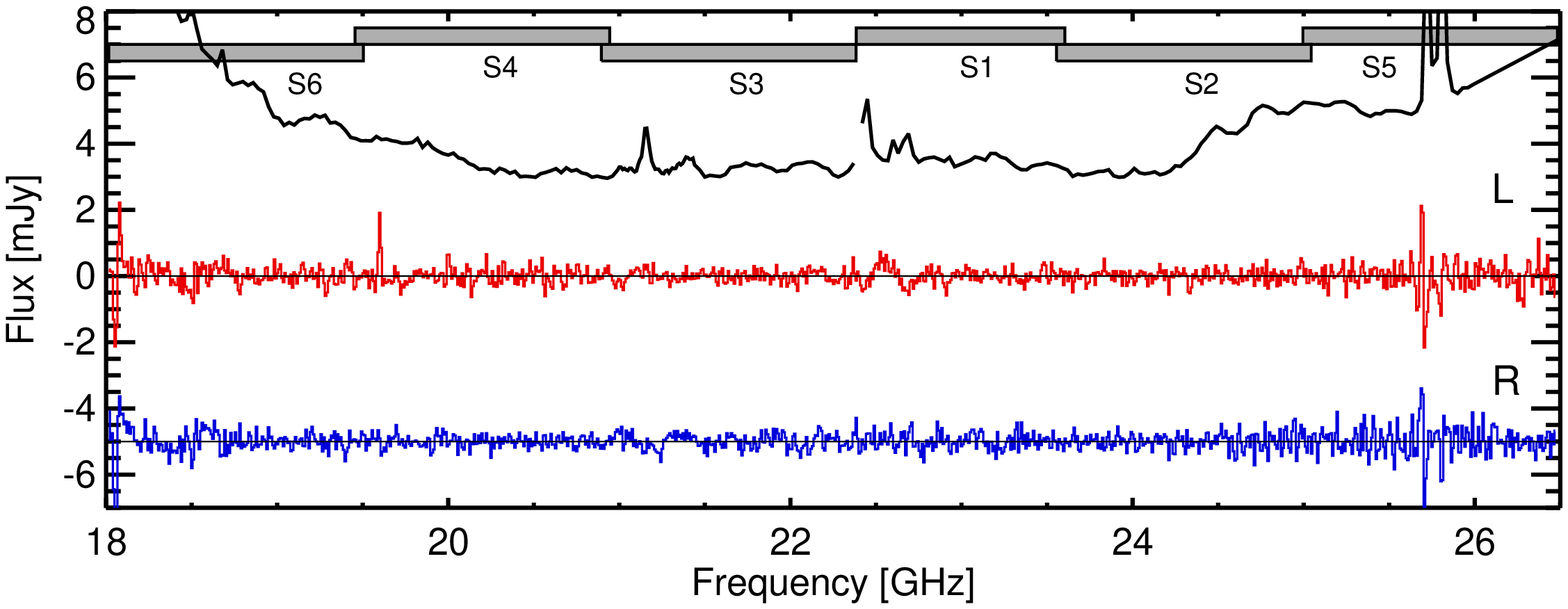,width=6.8in}

\epsfig{file=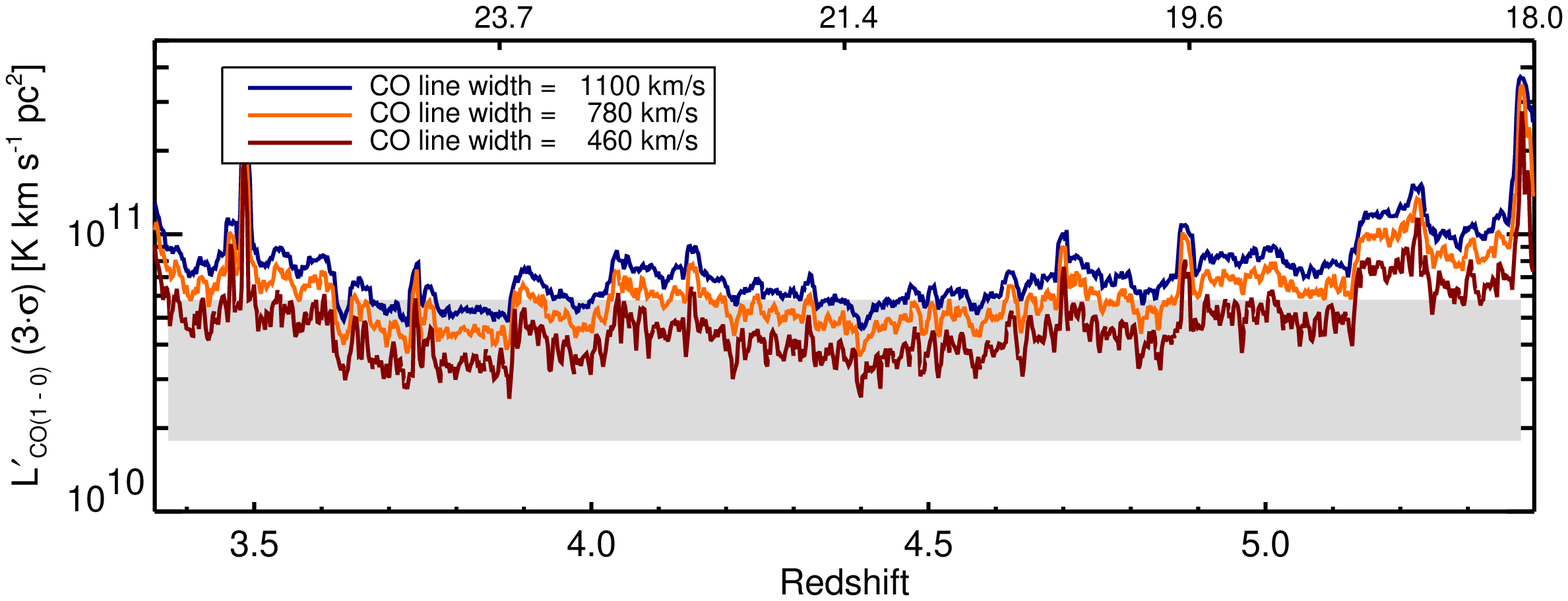,width=6.8in}
\caption[K-band spectra of HDF850.1 and corresponding limits to the 
\co line luminosity.]{\textbf{a)} K-band spectra of HDF850.1 produced from 
the best 47\% of the data, or 12.3 and 13.9 hours of on-source integration 
time included in the left and right polarization spectra, respectively. 
 Channel widths are 8.786~MHz. The black solid-line shows
 the scaled K-band receiver temperature sampled at 30~MHz resolution. 
Receiver resonance lines appear at various frequencies across the band. 
Grey boxes show the 6 frequency
 tunings (each $\sim$1.5~GHz wide) adopted to cover the full K-band window.
 Receiver resonances, such as the one at 25.7~GHz, also appear in the
 HDF850.1 spectra.
 Another noteworthy feature is the emission line at 19.6~GHz seen in the left
 polarization spectrum 
but not in the receiver temperature curve. This emission feature is also 
observed in the calibration spectra of 3C295, so we believe it to 
be another 
receiver resonance line. \textbf{b)} The 3-$\sigma$ upper-limits to the 
\co line luminosity calculated from the co-added left and right polarization
 spectra,
 assuming CO line widths of 1100, 780 and 460~km~s$^{-1}$. Also shown as the 
shaded region, is the range encompassed by the first
 and third quartiles of the CO line luminosities in the first 12 SMGs detected 
 in \hij CO line emission (see Greve et al.\ 2005). 
}
\label{fig:kband}
\end{figure*}

\begin{itemize}

\item Weather variations over short timescales (on order of the length of 
 a single scan of 1 minute duration) may result in inaccurate 
subtraction of the atmospheric 
 contribution to the system temperature across the band. This is particularly
 problematic at Q-band frequencies, where clouds passing overhead may lead to 
 rapid changes in the temperature of the atmosphere within a single beam. 

\item There may be interference from sources along the path of the analog 
signal, originating at the receiver on the telescope and travelling
 to the spectrometer backend in the GBT control room. An example of this 
is the 50~MHz ripple which 
appears in the left polarization K-band spectra taken during May, 
2005 (Figure~\ref{fig:rip50}), and
 is due to temperature variations in the equipment room where
the spectrometer is located. These variations 
cause standing waves in the connectors which manifest themselves as 
ripples in the spectral baseline.

\item Resonances in the receiver feeds may cause a loss of power at certain
 frequencies, leading to emission or absorption lines, sometimes referred 
to as `suck-outs'. When calculating $(\rm ON-OFF)/OFF$ from  
the on-source and off-source spectra, there will be a feature which is 
either in emission or absorption, depending on whether the power loss is 
in the on, or the off-beam. These `suck-outs' are apparent in the K-band  
receiver temperature curve shown in Figure~\ref{fig:kband}(a) (e.g. at 
frequencies of $\sim$22.6~GHz and $\sim$25.7~GHz), and may also be present
in the Q-band receiver system (however at present, no high-resolution receiver 
temperature data are available for this receiver).

\item A high-frequency ripple severely affected the right polarization 
Q-band spectra taken in December, 2005 (but may also be present in the 
left polarization at a lower amplitude), and is of an unknown origin. 

\end{itemize}

\noindent Data which was infected by any of these artefacts was not included 
when creating the final spectra. Thus $\sim$53$\%$ of the original data were
discarded.

\begin{figure*}
\epsfig{file=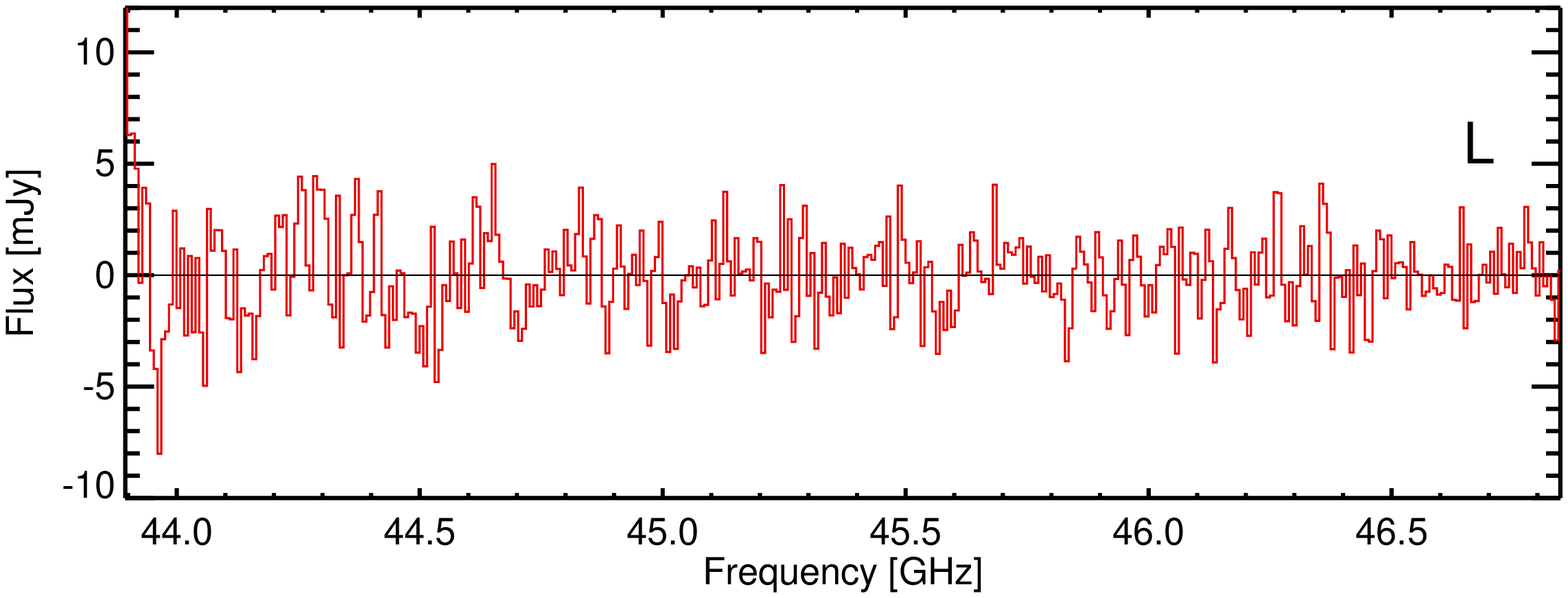,width=6.8in}

\epsfig{file=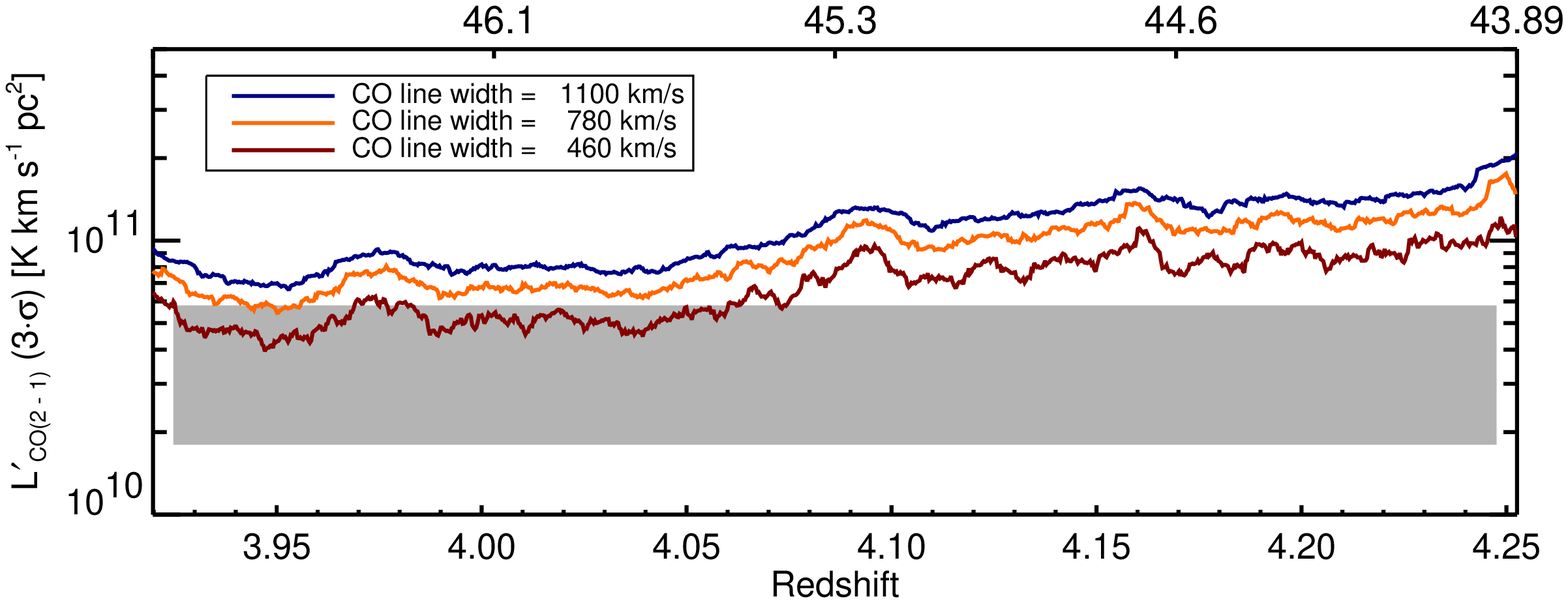,width=6.8in}
 \caption[Q-band spectra of HDF850.1 and corresponding limits to the 
\cotwo line luminosity.]
{\textbf{a)} The Q-band spectra of HDF850.1 showing only the 
best of the combined left polarization spectra sampled at 7.813~MHz resolution.
 The right polarization spectra 
suffered from a severe baseline ripple of unknown origin so are not included 
here. The redshift range (z~$\sim$~3.91 - 4.25) covered by this search for
 \cotwo improves 
upon the K-band spectra centered at ~22.5~GHz, where receiver resonance 
lines made it difficult to place constraints on the presence of \co line
 emission. \textbf{b)} The luminosity limits on \cotwo derived from the 
noise in the left polarization, Q-band spectrum, assuming CO linewidths of
 1100, 780 and 460~km~s$^{-1}$. As in Figure~\ref{fig:kband}(b), we also 
show the range defined by the first and third quartiles of the 
CO line luminosity 
(median $L'_{\rm CO} = 3.8\times 10^{10}$~K~km~s$^{-1}$~pc$^2$),
 as given in Greve et al.\ (2005).
}
\label{fig:qband}
\end{figure*}

\section{Results and Discussion}
\label{sec:results}

We do not find any evidence for \co line emission in our
 K-band spectra. Figure~\ref{fig:kband}(a) shows the final K-band spectra of 
HDF850.1 over the observed frequency range, 
$\nu_{\rm obs}$~=~18.0~to~26.5~GHz. 
Due to the presence of residual baseline features, $\sim$53$\%$ of the 
raw data are not included in the final, co-added spectra shown
 here. The most noticeable contaminant of our spectra are the 
receiver resonance features, appearing at various frequencies
along the K-band receiver temperature spectra (top curve, 
Figure~\ref{fig:kband}(a)). 
These resonances originate within the feed horns and 
their amplitudes depend strongly on the weather conditions. Although 
these features prevent us from placing any constraints on the presence  
of \co line emission over certain redshift intervals, 
 we are still able to obtain \co line luminosity limits over much of the 
K-band window.

\subsection{CO line luminosity limits}

We calculate 3-$\sigma$ upper limits to the \co and \cotwo line luminosities 
across the available K-band and Q-band spectra, respectively. We assume 
a range of CO line widths, $\Delta V_{\rm line}$~=~460, 780 and 
1100~km~s$^{-1}$, which represent the first, median and third quartiles 
of the linewidths in the first 
12 SMGs detected in \hij CO line emission (see Greve et al.\ 2005).  
The 3-$\sigma$ upper limit to the CO line integrated intensity is 
given by (e.g. Isaak, Chandler \& Carilli 2004),
 $3\cdot \sqrt{ \Delta V_{\rm line}/\Delta V_{\rm channel}}\cdot \sigma_{\rm channel} 
\cdot \Delta V_{\rm channel}$ 
(in units of Jy~km~s$^{-1}$), where the velocity width of a channel, 
$ \Delta V_{\rm channel}$, and the r.m.s. per channel, $\sigma_{\rm channel}$, 
both vary across the combined spectra. The limits to the integrated
 line intensity are 
converted to 3-$\sigma$ upper limits to the CO line luminosity, 
$L^{\prime}_{\rm CO}$, following the expression given by 
Solomon, Downes \& Radford (1992). These limits are shown in 
Figure~\ref{fig:kband}(b) for the \co line in the K-band spectra, and in 
Figure~\ref{fig:qband}(b) for the \cotwo 
line in the Q-band spectrum. For comparison, we also plot the median CO 
line luminosity measured in the first 12 SMGs detected in 
\hij ({\it J$_{\rm upper}$}~$\ge$~2) CO line emission (see Greve et al.\ 2005).
 
The fact that neither \co, or \cotwo line emission is detected in HDF850.1,
 can be explained by three possible scenarios; \textit{i)} the total 
molecular gas mass in this object is low, resulting in CO line 
emission that is weaker than our line luminosity limits, \textit{ii)} the
frequency of the emission line coincides with that of a receiver
resonance feature,
or \textit{iii)} the redshift of this object is such that the 
emission frequency of the \co line is outside the range accessible to 
the K-band receiver, a possibility which has a 10-14$\%$ probability 
according to the photometric redshift estimate of Aretxaga et al.\ (2003). 

Under the assumption that \textit{i)} is the correct scenario, 
the limits to the \co line luminosity can be used to estimate the limits on
 the total molecular gas mass, $M_{\rm H_2}$, 
in HDF850.1. Adopting the relationship:
 $M_{\rm H_2} = \alpha L'_{\rm CO(1-0)}$ 
($\alpha \sim1$~M$_{\odot}$(K~km~s$^{-1}$~pc$^2)^{-1}$), appropriate for 
 nearby ultraluminous infrared galaxies (Downes \& Solomon 1998), 
we can estimate an upper limit to the molecular gas mass contained within
 HDF850.1, under the assumption that its redshift is in the range, 
3.3~$\le z \le$~5.4. 
 The 3-$\sigma$ limit to the \co line luminosity is in the range,
 $L'_{\rm CO} \la (3.7 - 8.3)\times 10^{10}$~K~km~s$^{-1}$~pc$^2$, 
depending on the assumed line width and redshift. By first accounting for
a lensing amplification factor of 3 (Dunlop et al.\ 2004), these
 limits on the CO line luminosity translate 
directly to a molecular gas mass,
 $M_{\rm H_2} \la (1.2 - 2.8)\times 10^{10}$~M$_{\odot}$. The 
lensing amplification factor has been calculated for the FIR/submm 
emission region, which we are assuming is co-spatial with the \loj 
CO line emission region, resulting in an equal lensing factor.
 This assumption may not be valid, but without high angular 
resolution observations of both the FIR/submm and the 
\loj CO emission line regions, it remains uncertain.

\subsection{CO and Far-Infrared luminosities}

We can assess whether the CO line luminosity limits achieved here, are 
sufficient to have detected CO line emission over the observed redshift 
interval, by considering the CO line luminosity 
($L'_{\rm CO}$) predicted by the estimated far-infrared luminosity 
($L_{\rm FIR}$) of HDF850.1 within the context of the locally observed
 $L_{\rm FIR} - L'_{\rm CO}$ relation.    
In nearby galaxies there exists a well-established correlation  
between far-infrared luminosity and CO line luminosity 
(e.g. Young \& Scoville 1991), though it is unclear whether this 
relation truly arises from a direct 
dependance of star-formation rate (as traced by $L_{\rm FIR}$) on the total 
molecular gas mass 
(as traced by $L'_{\rm CO}$). Furthermore, this relationship 
appears to deviate from a power-law at high far-infrared luminosities 
($L_{\rm FIR} \ga 10^{12}$~L$_{\odot}$), which are characteristic of the SMG 
population. Despite the uncertainties in this relation, we converted the
 estimated far-infrared luminosity in HDF850.1 to 
an expected \co line luminosity, in order to determine if our limits on
 the \co and \cotwo line luminosities are sufficiently sensitive for us to 
have confidently expected a CO detection.

\begin{figure}
\epsfig{file=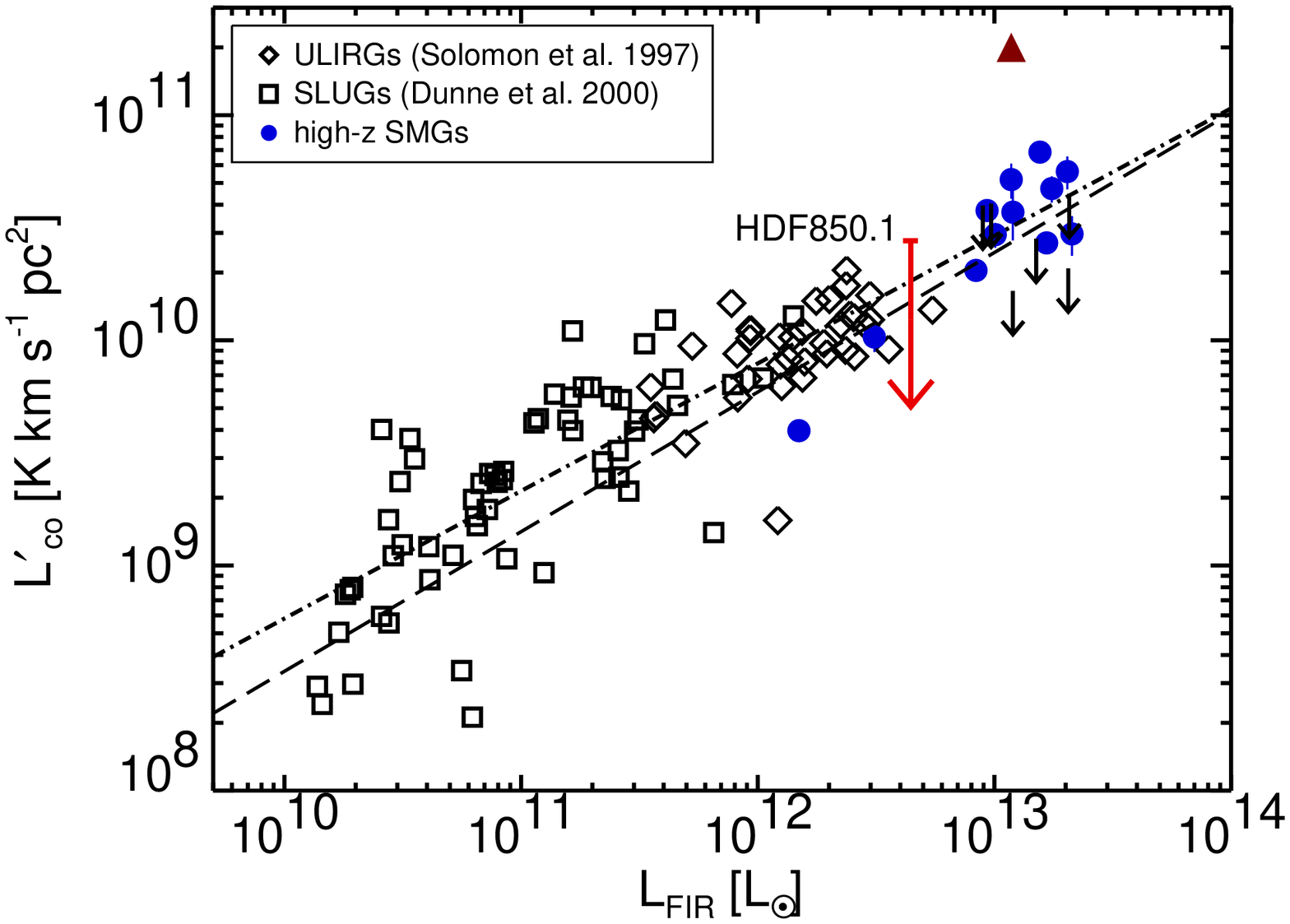,width=\hsize}
 \caption[CO line luminosity vs. far-infrared luminosity for low-redshift
LIRGs/ULIRGs and high-redshift SMGs.]{The relation between 
CO line luminosity and far-infrared luminosity in various 
samples of luminous infrared galaxies and AGN. The \textit{diamonds} 
are from the \co line observations of nearby ULIRGs presented in Solomon
et al.\ (1997), while the \textit{squares} are from the SLUGS sample of 
nearby LIRGs/ULIRGs observed at submm wavelengths by Dunne et al. (2000b). 
The \textit{solid circles} and \textit{short arrows} represent the sample 
of high-redshift SMGs which 
have been observed in \hij CO line emission (where both $L_{\rm FIR}$ and 
$L'_{\rm CO}$ has been corrected for lensing when applicable) and the 
\textit{solid triangle} 
is for the one SMG detected in \co line emission (Hainline et al.\ 2006).
The range in limits to the
CO line luminosity in HDF850.1 measured here (under the assumption of 
a redshift, 3.3~$\la z \la$~5.4) are shown as the \textit{long solid arrow}, 
and have been corrected for a lensing magnification factor of 3.
We show the fit presented in Greve et al.\ (2005) to a ULIRG+SMG sample
(\textit{dashed line}), along with the fit derived here to the low-redshift
objects plotted, and only those SMGs which are detected in CO line emission
(\textit{dot-dashed line}).  
}
\label{fig:lcolfir}
\end{figure}

Following Neri et al.\ (2003) and Greve et al.\ (2005), we calculate the 
far-infrared luminosity for HDF850.1 according to, 
$L_{\rm FIR} \sim 1.9\times 10 ^{12}~S_{850}\rm [mJy]$~L$_{\odot}$ 
(Blain et al.\ 2002), under the assumption of a modified greybody with 
dust temperature 
$T_{\rm d} = 40$~K, and emissivity index
 $\beta= 1.5$, where $S_{850}$ is the observed 850~$\mu$m flux density.
Although various measurements of the 850~$\mu$m flux density in HDF850.1 
exist in the literature, the differences are not significant and we 
adopt the original value presented by
 Hughes et al.\ (1998), $S_{850} = 7.0 \pm 0.5$~mJy. This leads to an 
estimated far-infrared luminosity, 
$L_{\rm FIR} \sim (13.3 \pm 1.0)m^{-1}\times 10 ^{12}$~L$_{\odot}$ (where $m$
is the magnification factor due to gravitational lensing, believed 
to be $\sim$3; Dunlop et al.\ 2004). 

In Figure~\ref{fig:lcolfir}, we compare the estimated limits on the 
$L_{\rm FIR} - L'_{\rm CO}$ parameter space obtained here for HDF850.1, with 
the $L_{\rm FIR} - L'_{\rm CO}$ relation observed in other SMGs and various 
low-redshift galaxy samples. The ultraluminous infrared galaxy (ULIRG) 
sample observed in \co by Solomon et al.\ (1997) is also included, along with 
the SLUGS sample of Dunne et al.\ (2000b) with \co line 
luminosities taken from the literature (Sanders et al.\ 1985, 1986, 1991;
Young et al.\ 1995; Casoli et al.\ 1996; Chini, Kr\"{u}gel \& Lemke 1996;
Maiolino et al.\ 1997; Solomon et al.\ 1997; Lavezzi \& Dickey 1998).
 More recent measurements of the nuclear \co and \cothree line
emission in a subset of the SLUGS are presented in Yao et al.\ (2003). 
The SMG sample consists of those objects in which searches have been
conducted for \hij CO line emission (Frayer et al.\ 1998, 1999; 
Neri et al.\ 2003; 
Genzel et al.\ 2003; Sheth et al.\ 2004; Greve et al.\ 2005; 
Kneib et al.\ 2005; Tacconi et al.\ 2006), and the 
one object, SMM~J13120+4242 at z=3.4, for which the \co line has been 
detected (Hainline et al.\ 2006). 
The far-infrared luminosities
of these objects are calculated following the same prescription 
as that adopted for HDF850.1, with submm/mm 
 flux densities taken from the literature (Smail et al.\ 1997, 1998; 
Ivison et al.\ 1998; Barger, Cowie, \& Sanders 1999; Dey et al.\ 1999; 
Cowie, Barger, \& Kneib 2002; 
Scott et al.\ 2002; Chapman et al.\ 2003b, 2005; 
Greve et al.\ 2004).
 For the purpose of 
comparison with the nearby galaxies detected in \co, we follow
Greve et al.\ (2005) and assume that for the \hij CO lines 
observed in SMGs, 
$ L'_{\rm CO(4-3)} / L'_{\rm CO(1-0)} = L'_{\rm CO(3-2)} / L'_{\rm CO(1-0)} = 
L'_{\rm CO(2-1)} / L'_{\rm CO(1-0)} = 1$, 
 corresponding to optically-thick, thermalized CO emission. These
data are plotted in Figure~\ref{fig:lcolfir}, with appropriate 
corrections applied to both $L_{\rm FIR}$ and $L'_{\rm CO}$ to account
 for magnification by gravitational lensing in the 6 SMGs believed to be
 lensed (assuming co-spatial far-infrared and CO emission line regions).

Also plotted in Figure~\ref{fig:lcolfir} is a fit to the 
$L_{\rm FIR} - L'_{\rm CO}$ relation
derived from a combined sample of ULIRGs and SMGs by
 Greve et al.\ (2005) where
$\log{L'_{\rm CO}} = (0.62\pm0.08)\log{L_{\rm FIR}} + (2.33\pm0.93)$. 
We find a similar relation when we fit to the luminosities in the
combined low-redshift LIRGs/ULIRGs and SMG sample plotted here,
$\log{L'_{\rm CO}} = (0.57\pm0.03)\log{L_{\rm FIR}} + (3.10\pm0.34)$.
These fits may be used to compare the estimated far-infrared luminosity in 
HDF850.1 to the measured CO line luminosity limits. After correcting
for amplification by gravitational lensing, the estimated far-infrared
 luminosity in HDF850.1 makes it one of the least intrinsically 
luminous SMGs that has been searched for CO line emission,
 as most in the sample %(11/14)
are believed to be unlensed (see Greve et al.\ 2005). Adopting the 
 estimate for $L_{\rm FIR}$ in HDF850.1, the fit to the 
$L_{\rm FIR} - L'_{\rm CO}$ relation by Greve et al.\ (2005) would predict,
 $L'_{\rm CO} = 2.9m^{-1}\times 10^{10}$~K~km~s$^{-1}$~pc$^2$, while the
fit presented here would predict,
 $L'_{\rm CO} = 3.4m^{-1}\times 10^{10}$~K~km~s$^{-1}$~pc$^2$. These values 
are generally lower than the 3-$\sigma$ CO line luminosity limits
 achieved in our K-band and Q-band spectra.

 We are unable to draw any conclusions as to the
validity of the photometric redshift technique applied to HDF850.1.
 Although we have removed $\sim$53\% of the 
original K-band data due to various spectral baseline irregularities,
 a further $\sqrt{2}$ decrease in the noise would still not be 
sufficient to obtain an $L'_{CO}$ limit that was significantly 
inconsistent with the expected CO content, given the uncertainties and
 the dispersion in the estimated gas masses of SMGs with CO detections.
The \co and \cotwo line luminosity limits presented here are not of 
sufficient depth to exclude the presence of CO line emission within
the redshift interval, 3.3~$\la z \la$~5.4.

\section{Conclusions}

We present a broadband, GBT spectroscopic search for \co and \cotwo
 line emission in the high-redshift SMG, HDF850.1 using the K-band 
(18.0 to 26.5~GHz) and Q-band (40.0 to 48.0~GHz) receivers.
 Although we do not detect any CO line 
emission in this object, our constraints on the CO line luminosity are 
approaching that predicted by the far-infrared luminosity within
the context of the local $L_{\rm FIR} - L'_{\rm CO}$ relation. These 
GBT results are still consistent with HDF850.1 lying in the redshift 
interval,  3.3~$\la z \la$~5.4, based on our previous rest-frame 
far-infrared-to-radio photometric redshift estimate.

The GBT has recently been successful in detecting the \co line 
in 3 quasar host galaxies; APM~08279+5255 at z=3.9, 
PSS~J2322+1944 at z=4.1 and 
BR~1202-0725 at z=4.7 (Riechers et al.\ 2006), while Hainline
 et al.\ (2006) present a detection of \co line emission in the SMG 
SMM~J13120+4242 at z=3.4. These 4 objects detected in \co line emission 
with the GBT, were previously known to exhibit strong \hij CO line-emission, 
and therefore to contain large masses of warm molecular gas.  
 With these prior \hij CO detections, the redshifts for the molecular
 emission-line regions were constrained to $\la$100~km~s$^{-1}$
 enabling a more efficient use (i.e. a narrower frequency search) of 
their available GBT observing time.
 Given the prior uncertainty in both the redshift and CO line intensity 
of HDF850.1, our experiment is quite distinct from the `tuned' GBT 
observations described above.  
This is the first broad-bandwidth cm-wavelength search for CO-line emission 
in a high-redshift object (guided by a radio-to-FIR photometric redshift),
with no previous detections of molecular line emission or an optical redshift.

Considering future possibilities, 
the gaseous medium within SMGs is expected and perhaps already shown
to be warm and dense, 
 and hence the \hij CO line transitions should be more intense than the
 {\it J}=1-0 transition (Weiss et al.\ 2005). At z$\sim$$2-4$ (typical 
of the SMG population) the \hij ({\it J}~$\ge$~2) CO transitions are 
redshifted into the $\sim$$70- 310$~GHz atmospheric windows. Thus we are 
optimistic that CO line searches, using broadband mm-wavelength receivers 
on sensitive facilities such as the LMT, PdBI or CARMA, will 
be more successful in obtaining unambiguous spectroscopic 
redshifts for the optically obscured SMG population of starburst galaxies.

\section*{Acknowledgments}

We are very grateful to the entire Green Bank staff for their help and 
patience throughout the course of these observations. In particular, 
we would like to thank Carl Bignell, Ron Maddalena, Dana Balser, Karen O'Neil,
Tony Minter, Frank Ghigo, Glen Langston, Brian Mason, Jay Lockman, Phil Jewell 
and Richard Prestage. J.W. would like to thank Paul Kondratko for informative 
discussions on GBT data reduction. J.W. thanks the Department of Astrophysics
 at INAOE for a graduate student scholarship and the SAO for the funding 
provided by 
a predoctoral student fellowship. D.H.H., J.W. and I.A. are supported by
 CONACYT grant 39953-F. This work is partially funded by CONACYT grant 39548-F.
We thank the anonymous referee for helpful suggestions.

\end{document}